\documentclass{ws-p8-50x6-00}
\def\RMP{\em Rev. Mod. Phys.}
\def\EPJC{{\em Eur. Phys. J.} C}

\def\PRC{{\em Phys. Rev.} C}
\def\htm{\hat{t}}
\def\la{\langle}
\def\ra{\rangle}
\def\kav{\la k_T \ra}
\def\k2av{\la k_T^2 \ra}
\def\knav{\la k_T^2 \ra}
\def\be{\begin{eqnarray}}
\def\ee{\end{eqnarray}}
\newcommand{\f}[2]{\frac{#1}{#2}}
\begin{document}

\title{NUCLEAR EFFECTS IN HIGH ENERGY PHOTON AND NEUTRAL PION PRODUCTION}

\author{G\'abor Papp$^{1,2}$, George Fai$^{1}$ \lowercase{and}
 P\'eter L\'evai$^{3}$}
\address{$^{1}$CNR, Department of Physics, 
Kent State University, Kent, OH 44242, USA\\
$^{2}$ HAS Research Group for Theoretical Physics, E\"otv\"os 
University, Budapest\\
$^{3}$ KFKI RMKI, P.O. Box 49, Budapest, 1525, Hungary} 
%
%
%
\maketitle

\abstracts{We study the mechanism of the $\gamma$ and $\pi^0$
enhancement at large transverse momenta in proton-nucleus collisions. A
proposed mechanism suggests an increase in the width of the transverse
momentum distribution of partons in a semihard process.}

\vspace*{3mm}
The nuclear enhancement of particle production, known as
the Cronin effect,~\cite{Cronin} received recent experimental
attention by E706,~\cite{E706} studying the high $p_T$ pion and direct
photon production.
In this talk we concentrate on direct photon and neutral pion
production and propose a new physical mechanism responsible for the
increase of the production rate in nuclear targets.

Parton cross sections are calculable in pQCD at high energy to leading
order (LO) or next-to-leading-order (NLO).\cite{OptimQres} The parton
distribution functions (PDFs) and fragmentation functions (FFs),
however, need to be fitted to data. In recent NLO calculations the
various scales are optimized.~\cite{OptimQres} Alternatively, an
additional non-perturbative parameter, the width of the {\it intrinsic
transverse momentum} ($k_T$) distribution of the partons is
introduced.~\cite{Huston95} We choose the latter method and use $k_T$
phenomenologically, expecting that its importance
will decrease in higher orders of pQCD.

\vspace*{2mm}
In the lowest-order pQCD-improved parton model,
direct pion production can be described in $pA$ collisions by
\be
\label{fullpi}
  E_{\pi}\f{d\sigma_\pi}{d^3p} &=&
        \sum_{abcd}\!\int\!d^2b\,t_A(b) \int\!dx_{1,2}
	d^2k_{T_{1,2}}\,g_1(\vec{k}_{T_1},b) g_2(\vec{k}_{T_1}) \\
	&& f_{1}(x_1,Q^2)\
        f_{2}(x_2,Q^2)\  K  \f{d\sigma}{d\htm}\,
   \frac{D(z_c,{\widehat Q}^2)}{\pi z_c} \ ,\nonumber
\ee
where $f_{1}(x,Q^2)$ and $f_{2}(x,Q^2)$ are the PDFs of partons $a$ and
$b$, and $\sigma$ is the LO cross section of the appropriate partonic
subprocess $(ab\to cd)$. The K-factor accounts for higher order
corrections.~\cite{Owens87} Comparing LO and NLO calculations a nearly
constant value, $K\approx 2$, is obtained as a good approximation of the
higher order contributions in the $p_T$ region of
interest.~\cite{Wong98} In eq.(\ref{fullpi}) $D(z_c,{\widehat Q}^2)$ is
the FF of the pion, with ${\widehat Q} = p_T /z_c$, where $z_c$ is the
momentum fraction of the final hadron.  We use NLO parameterizations of
the PDFs\cite{MRST98} and FFs~\cite{BKK} with fixed scales and the
partons are assumed to have transverse momenta described by the
distribution functions $g_i(\vec{k})$. The nuclear effects are
hidden in 
the impact parameter dependence of this distribution and in the
nuclear thickness function $t_A(b)=\int\!dz\,\varrho(\vec{b},z)$. Here
$\varrho(\vec{r})$ is the density distribution of the nucleus.
Direct $\gamma$ production is described similarly.

In this talk we discuss two idealized density distributions, the
homogenous (sharp sphere) nucleus and a Gaussian density
profile. The transverse
momentum distribution $g({\vec k}_T)$, is parameterized to be Gaussian, 
$\exp(-k_T^2/\langle k_T^2 \rangle)/{\pi \langle k_T^2 \rangle}$ with
a 2-dimensional width of the $k_T$
distribution, $\langle k_T^2 \rangle$, related to the average $k_T$ of
one parton 
as $\langle k_T^2 \rangle = 4 \langle k_T \rangle^2 /\pi$. Applying the 
model to data from $pp\rightarrow \pi^0  X$  and $pp\rightarrow \gamma
X$ experiments we deduced
$\kav_{pp}^\gamma=0.545\log{\sqrt{s}}-0.9$ and
$\kav_{pp}^\pi=0.459\log{\sqrt{s}}+0.092$ in the
energy range $\sqrt{s}=20$--$65$ GeV.~\cite{plf99}

The standard physical explanation of the nuclear enhancement
(Cronin effect) is that the proton traveling
through the nucleus gains extra transverse momentum due to 
random soft collisions and the partons enter the final hard
process with this extra  $k_T$.\cite{Wang9798} 
We write  the width of the transverse momentum distribution of the partons
in the incoming proton as
\begin{equation}
\label{ktbroadpA}
\knav_{pA} = \knav_{pp} + C \cdot h_{pA}(b) \ . 
\end{equation}
Here $h_{pA}(b)$ is the number of {\it effective} 
nucleon-nucleon collisions at impact parameter $b$ 
imparting an average transverse momentum squared $C$. 
Naively all possible soft interactions are included,
but such a picture leads to a target-dependent $C$.\cite{plf99}
A more satisfactory description is obtained with a ``saturated''
$h_{pA}$, where it takes
at most one semi-hard ($Q^2 \sim 1$ GeV$^2$) collision
for the incoming proton to loose coherence, 
resulting in an increase of the width of its $k_T$
distribution.~\cite{plf99} 
This is approximated by a smoothed step
function with a maximum value of unity.
The saturated Cronin factor is denoted by  $ C^{sat}$.

{\bf i. Sharp sphere nucleus}: The thickness function of a sharp sphere
nucleus is $t_A(b) = 2 \rho_0 \sqrt{R_A^2-b^2}$
with $\rho_0=0.16$ fm$^{-3}$. Calculations show that if all possible 
soft collisions are included, the momentum square imparted per collision is
target dependent,
$C^{all}_{pBe}=0.8\pm 0.2$, $C^{all}_{pTi}=0.4\pm 0.2$ and
$C^{all}_{pW}=0.3\pm 0.2$ GeV$^2$.  In the saturating model an opposite
tendency may be observed with best fits being $C^{sat}_{pBe}=0.7$, 
$C^{sat}_{pTi}=0.85$ and $C^{sat}_{pW}=1.35$ GeV$^2$, respectively,
however, with much larger tolerance, allowing for a common value,
$C^{sat}=1.1-1.2$ GeV$^2$ to be used in the $p A \rightarrow \pi^0 X$
experiment with $A=Be$, $Ti$ and $W$.\cite{Cronin} Surprisingly, the
same value of $C^{sat}$ describes the $\gamma$ production at
$\sqrt{s} \approx 30-40$ GeV as well. 

{\bf ii. Gaussian nucleus}: The thickness function of a Gaussian density
distribution is $t_A(b) = \varrho_0 e^{-\alpha b^2} \sqrt{\pi/\alpha}$
with $\alpha=\pi (\varrho_0/A)^{2/3}$. For the $Be$ target the maximum
value of $h_{pA}$ is below 2 in this case, hence the soft collision and
the saturated descriptions are identical. Calculations show that
$C^{all}_{pBe}=1.3\pm 0.2$, $C^{all}_{pTi}=0.72\pm 0.1$ and
$C^{all}_{pW}=0.65\pm 0.1$ GeV$^2$. Once again, for the saturated model
we have $C^{sat}_{pBe}=1.3$, 
$C^{sat}_{pTi}=1.15$ and $C^{sat}_{pW}=1.55$ GeV$^2$, respectively, with
large uncertainties. The
common value of $C^{sat}=1.4$ GeV$^2$ gives a good description of the
experiment.


We interpret $C^{sat}$ as the square of the typical transverse
momentum imparted in {\it one} semi-hard collision prior to the hard
scattering. Both descriptions could be made consistent with the data on
the $Ti$ and $W$ targets, but the soft-collision model is ruled out if
we include the whole range of $A$ from $Be$ to $W$.
The proposed mechanism gives better agreement with
experimental data, although the extracted value of
$C^{sat}$ is slightly target dependent, showing an increase at the heaviest
nucleus. This may indicate a contribution from soft collisions. More
precise (NLO) calculations are required to analyze this effect.
This picture and the energy dependence of $C^{sat}$ need 
to be tested as functions of $A$ at different energies.

\section*{Acknowledgments}
This work is supported by the US-DOE grant, DE-FG02-86ER40251, and by
Hungarian grant OTKA, F019689.


\end{document}